\documentstyle[12pt,epsf]{article}
\renewcommand{\bar}[1]{\overline{#1}}

\newcommand{\gsim} {\buildrel > \over {_\sim}}

\def\ru1{\rule[-0.4truecm]{0mm}{1truecm}}

\begin{document}

\begin{flushright}
{\footnotesize SLAC-PUB-7329\\
USM-TH-71}\\
hep-ph/97xxxx
\end{flushright}

\bigskip\bigskip
{\centerline{\Large \bf
Polarization  Asymmetry Zero in Heavy Quark}
{\centerline{\Large \bf
Photoproduction and
Leptoproduction\footnote{\baselineskip=13pt Work partially supported by the
Department of Energy, contract DE--AC03--76SF00515, by Fondecyt (Chile) under
grant 1960536, and by a C\'atedra Presidencial (Chile).}}}

\vspace{22pt}

\centerline{
\bf Stanley J. Brodsky$^a$ and Ivan Schmidt$^{b}$}

\vspace{8pt}
{\centerline{$^a$Stanford Linear Accelerator Center,}}

{\centerline{Stanford University, Stanford, California 94309, USA}}


\vspace{8pt}
{\centerline {$^{b}$Departamento de F\'\i sica, Universidad T\'ecnica
Federico
Santa Mar\'\i a,}}

{\centerline {Casilla 110-V, 
Valpara\'\i so, Chile}}



\vspace{10pt}
\begin{center} {\large \bf Abstract}

\end{center}

\baselineskip12pt
We demonstrate two novel features of the sea-quark
contributions to the polarized structure functions and photoproduction cross
sections, a zero sum rule and a zero crossing point of the polarization
asymmetry, which can be traced directly to the dynamics of the perturbative
tree-graph gluon-splitting contributions.   In particular, we show that the Born
contribution of massive quarks arising from  photon-gluon fusion gives zero
contribution to the logarithmic integral over the polarization asymmetry
$\int {d\nu \over \nu}\Delta
\sigma(\nu,Q^2)$ for any photon virtuality. The vanishing of this integral in
the Bjorken scaling limit then implies a zero gluon-splitting Born contribution
to the Gourdin-Ellis-Jaffe sum rule for polarized structure functions from
massive sea quarks. The vanishing of the polarization asymmetry at or near the
canonical position predicted by perturbative QCD  provides an important tool for
verifying the dominance of the photon-gluon fusion contribution to charm
photoproduction and for validating the effectiveness of this process as a
measure of the gluon polarization
$\Delta G(x,Q^2)$ in the nucleon. The displacement
of the asymmetry zero from its canonical position
is sensitive to the virtuality of
the gluon in the photon-gluon fusion subprocess, and it can provide a measure of
intrinsic and higher-order sea quark contributions.

\vfill
\centerline{
}
\vfill

\centerline{Submitted to Physics Letters B.}
\vfill
\newpage

\baselineskip 23.5pt

Deep-inelastic polarized lepton-polarized nucleon
scattering provides a unique testing ground for quantum chromodynamics,
challenging our theoretical understanding of hadron
structure\cite{Ansel}.
The Bjorken sum rule\cite{Bj}
for the non-singlet polarized nucleon structure functions gives a remarkable
connection of deep inelastic scattering to the weak axial coupling $g_A$. The
prediction that the QCD radiative corrections to the Bjorken sum rule are the
inverse of the radiative corrections to  the ${e^+ e^-
\to {\rm hadrons}}$ cross section at their commensurate moemnetum transfer
and energy scales provides a fundamental test of perturbative QCD\cite{CSR}.
There are also many important non-perturbative QCD aspects of the polarized
structure functions such as the contributions to nucleon spin from valence and
sea quarks\cite{SMC97},  the role of polarized
gluons arising through the axial coupling anomaly,  the influence of  quark
orbital angular momentum, and the apparently large and negative contribution
from the strange sea. It is even possible that the $s$ and
$\bar s$ momentum and helicity distributions are distinctly different in the
nucleon\cite{Ma}.

In this paper we will demonstrate two unexpected features of the sea-quark
contributions to photoproduction and leptoproduction cross sections:
a zero sum rule and a zero
crossing point of the longitudinal polarization asymmetry, which can be traced
directly to the dynamics of the perturbative tree-graph gluon-splitting
contributions.  The vanishing of the
polarization asymmetry at or near the canonical position predicted by
perturbative QCD provides an important tool for verifying the dominance of the
photon-gluon fusion contribution to charm photoproduction for real or virtual
photons and for validating the effectiveness of this process as a measure of the
gluon polarization
$\Delta G(x,Q^2)$ in the nucleon. The displacement
of the asymmetry zero from its canonical position
is sensitive to the virtuality of
the gluon in the photon-gluon fusion subprocess, and it can provide a measure of
intrinsic and higher-order sea quark contributions to the polarized structure
functions.

It is well known from the work of Burkert and Ioffe\cite{Ioffe}
that the polarized deep-inelastic
structure function $g_1(x,Q^2)$ is an analytic extension into the Bjorken
scaling region of the polarized photo-absorption cross section $\Delta
\sigma(\nu,Q^2)$, the same cross section difference that appears for real
photons in the Drell-Hearn-Gerasimov (DHG) sum rule\cite{DHG}.
The
sum rule we discuss here follows from a set of superconvergence relations
for the
DHG integral\cite{BS} and provides  new insights into the role of polarized
gluons and the gluon anomaly contribution from light sea quarks to the
Gourdin-Ellis-Jaffe (GEJ) singlet sum rule\cite{GEJ}.  In
particular, we shall show that the Born contribution of massive quarks arising
from gluon splitting gives zero contribution to the logarithmic integral over
the polarization asymmetry
$\int {d\nu \over \nu}\Delta
\sigma(\nu,Q^2)$ for any photon virtuality. The vanishing of this integral then
implies in the Bjorken scaling limit a zero gluon-splitting Born contribution to
the Gourdin-Ellis-Jaffe sum rule $\int^1_0 dx g_1^p(x,Q^2)$ from
massive sea quarks, in agreement with the triangle graph calculation
of Carlitz, Collins, and Mueller\cite{Carlitz}. It also gives new insights
into the factorization
scheme-dependence of the interpretation of the gluon anomaly contribution for
massless sea quarks emphasized by  Bodwin and Qiu\cite{Geoff}.

The diagrammatic analysis of polarized deep-inelastic scattering starts with the
virtual photon-proton polarized asymmetry ${\cal A}^{\gamma^*N}$\cite{Geoff}.
According to the factorization theorem, this can be separated into
the convolution of hard and soft pieces:
\begin{equation}
{\cal A}^{\gamma ^{\ast }N}(x,Q^{2})=
{\cal A}_h^{\gamma^{\ast}g}\left(x,\frac {Q^{2}}{\mu^{2}}\right)
\otimes \Delta G(x,\mu ^{2})
+{\cal A}_h^{\gamma ^{\ast}q}\left(x, \frac{Q^{2}}{\mu^{2}}\right)
\otimes\Delta q(x,\mu ^{2}),
\label{AA}
\end{equation}
where $\mu^2$ is the factorization
scale, $\Delta G$ and $\Delta q$ are the polarized gluon and quark distributions
respectively, and ${\cal A}_h^{\gamma^*g}$ (${\cal A}_h^{\gamma^*q}$) is the
hard part of the
polarized photon-gluon (quark) asymmetry. To be more precise, the asymmetries
${\cal A}$ are the differences of imaginary parts of forward
current-current matrix elements when the photon has the same or
opposite helicities to that of the target\cite{footnote}.

Although the polarization asymmetry ${\cal A}^{\gamma^*N}$ is a physical
quantity, the identification of the individual contributions given in Eq.
(\ref{AA}) have an intrinsic ambiguity for light mass sea quarks related to the
precise prescription by which the hard parts are defined.\cite{Carlitz}
Two general regularization
schemes have been proposed in the literature:\cite{Cheng}
the gauge-invariant scheme, which breaks chiral
symmetry, and the chiral-invariant scheme, which breaks gauge
invariance. In the gauge-invariant scheme, hard gluons do not contribute to the
first moment of $g_1^p(x)$\cite{Geoff}.  Since the anomaly corresponds to the
quantum breaking of the chiral symmetry which allows for the creation of pairs
of definite chirality even in the zero mass limit, a net sea-quark polarization
is expected\cite{Forte}. In this case the
anomaly is then identified with the sea quark
helicity density inside a gluon.

The chiral-invariant  scheme
is implemented by imposing a sharp perpendicular quark
momentum $k_{\perp}$ cutoff to separate hard and soft gluon momenta in the
photon-gluon asymmetry diagram calculation (see Fig. 1). In this case, the
effect of the anomaly from light sea quarks is shifted from the
helicity-dependent quark distribution to the hard gluon asymmetry. Then the
integral of this hard part is the anomaly, independent of the infrared regulator
that has been used to regulate collinear divergences, and the first moment of
$g_1^p(x,Q^2)$ contains an anomalous gluon contribution $-(\alpha_s /2 \pi)
\Delta G$.

In this paper we shall utilize a ``light-cone factorization scheme" which is
similar to the chiral-invariant scheme, but has a physical,
frame-independent parton model interpretation: the quark distributions
$q(x,\mu_{\rm fact})$ are defined from the absolute square of the
light-cone Fock state wavefunctions integrated up to
invariant mass ${\cal M} < \mu_{\rm fact}$ \cite{BL}. This can also be
interpreted as a cutoff in parton virtuality.
For example, a gluon with light-cone momentum fraction $x=k^+/P^+$ in a Fock
state of the proton with invariant mass
${\cal M}$ has Feynman virtuality $k^2 = x(M_p^2-{\cal M}^2)$.   Unlike
the sharp perpendicular quark momentum
$k_{\perp}$ cutoff, this definition of the parton distributions is
invariant under Lorentz boosts. The light-cone factorization scheme is however
gauge-dependent since the light-cone Fock wavefunctions describe particles
defined in the physical $A^+=0$ light-cone gauge.

\vspace{0.5cm}
\begin{figure}[htb]
\begin{center}
\leavevmode {\epsfysize=6.8cm \epsfbox{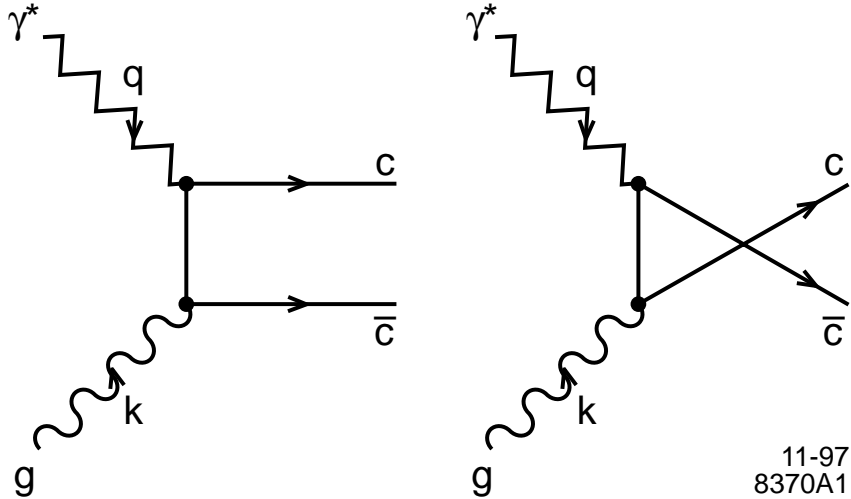}}
\end{center}
\caption[*]{\baselineskip 13pt
Photon-gluon fusion diagrams.}
\label{8370A01}
\end{figure}

Let us examine in more detail polarized inclusive $c \bar{c}$
pair production in deep inelastic scattering.
The dominant contribution to the partonic
subprocess at any photon virtuality is expected to derive from the photon-gluon
fusion diagrams
$\gamma^*g \to c
\bar c$ shown in Fig. 1.  The polarized
gluon distribution then contributes to the helicity-dependent charm structure
structure functions at order
$\alpha_s$, in accordance with the first term of Eq. (\ref{AA}).
Charm photoproduction is the
basis of proposed experiments at SLAC\cite{SLAC} and
HERMES\cite{HERMES}, because of its sensitivity to the polarized gluon
distribution.

Let us consider the logarithmic integral of the polarized virtual
photoabsorption
cross section:
\begin{equation}
\int _{0}^{x_{\max}}{\cal A}^{\gamma ^{\ast }g}(x)\ dx\propto \ \int
_{s_{th}}^{\infty }{ds\over s+Q^{2}}\big[\Delta \sigma (s,Q^2)\big],
\label{BB}
\end{equation}
where
$\Delta\sigma(s)=\sigma_P-\sigma_A$ is the difference of cross sections for
parallel and antiparallel photon-gluon helicities. Here
$x_{\max}=Q^{2}/(Q^{2}-k^{2}+4m^{2})$, $s_{th}=4m^{2}$, $s=(q+k)^2$, and
$q^2=-Q^2 (\leq 0)$ and $k^2 (\leq 0)$ are the virtualities of the photon and
gluon respectively.
In fact, the Born contribution to the integral is zero when the gluon is
on
its mass shell $k^2=0$, but for any photon virtuality $q^2$. To
prove this,  consider the process $\gamma a \to b c$, where $a, b$, and
$c$ are arbitrary fields of a renormalizable theory (as long as $a$ carries
nonzero spin). Then one can show from the absence of an  anomalous
moment of particle $a$ at lowest order, that the logarithmic integral of the
helicity-dependent part of the photoabsorption cross section must vanish in Born
approximation, i.e.
\begin{equation}
\int _{\nu _{th}}^{\infty }{d\nu \over \nu }\ \Delta\sigma _{\rm Born}(\nu
)\ =\ 0,
\label{CC}
\end{equation}
where $\nu$ is the photon laboratory energy\cite{BS}.  This is a
remarkable consequence of the Drell-Hearn-Gerasimov sum rule, reflecting the
canonical couplings of the fundamental particles in gauge theory. One of its
most interesting applications is a novel method for measuring the $W$ magnetic
and quadrupole moments to high precision in electron-photon
collisions\cite{BRS}.

In our application of this classical sum rule to polarized deep inelastic
scattering, the virtual photon plays the role of the target, and the gluon plays
the role of the on-shell photon. Since the photon is spacelike,
we change variables from $\nu$ to $s=(p+q)^2$, and
analytically continue to negative $q^2$.   We then obtain:
\begin{equation}
\int _{s_{th}}^{\infty
}{ds\over s+Q^{2}}\ \Delta \sigma _{\rm Born}(s,Q^2)\ =\ 0,
\label{DD}
\end{equation}
which is exactly the Born heavy quark integral of equation (\ref{BB}).
Thus we have shown that the Born heavy-quark integral vanishes not only in the
scaling limit, but for any $Q^2=-q^2$.

The DHG integral of the virtual photon-gluon fusion contribution for massless
gluons ($k^2=0)$ can be calculated explicitly, and in agreement with Eq.
(\ref{DD}),  the Born contribution  can be shown to vanish for any value of the
$Q^2$ and quark mass $m$.
The natural physics variable is the velocity $\beta$ of the quark in the
CM. If we define
$y=1-\beta^2=4m^2/s$ and $a=Q^2/4m^2$, the DHG integral reads:
\begin{equation}
\int_{0}^{1}{dy\over (1+ay)^{3}}\bigl\{ (1-ay)L+(3-ay)\beta \bigr\}=0,
\label{EE}
\end{equation}
where
$L=\log((1-\beta)/(1+\beta))$.

The vanishing of the DGH integral also implies that there must be a value of
$s=s_0$ and $\beta=\beta_0$ where  $\Delta \sigma (s_{0})=0$. In fact, the
the polarization asymmetry
will reverse its sign for any value of the ratio $a=Q^2/4m^2$.
In terms
of the Bjorken variable,
$x=Q^{2}/(s+Q^{2})=ay/(1+ay)$, the zero occurs in the range $0 < x < 1/2$ when
\begin{equation}
x_{0}={3\beta _{0}+L_{0}\over 4\beta _{0}+2 L_{0}}.
\label{FF}
\end{equation}
Figure 2 shows
the crossing point as a function of $Q^2/4m^2$ for zero gluon virtuality
$k^2=0$.

The vanishing of the polarization asymmetry at or near the canonical position
predicted by perturbative QCD can provide an important tool for verifying
the dominance of the photon-gluon fusion contribution to charm photoproduction
and for validating the effectiveness of this process as a measure of the gluon
polarization
$\Delta G(x,Q^2)$ in the nucleon.

\vspace{0.5cm}
\begin{figure}[htb]
\begin{center}
\leavevmode {\epsfysize=6.8cm \epsfbox{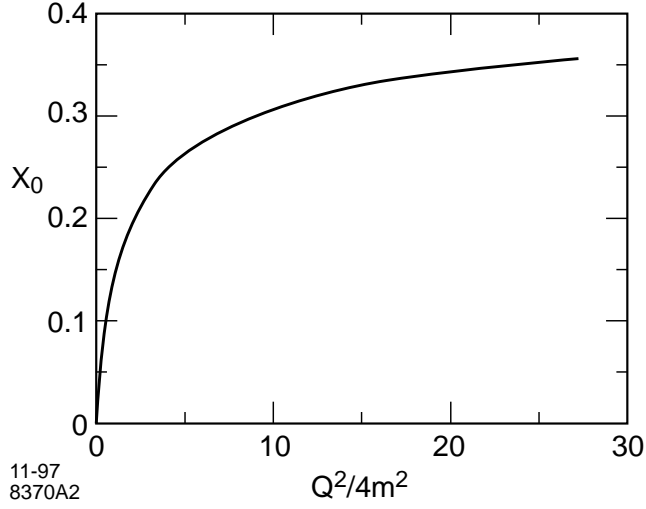}}
\end{center}
\caption[*]{\baselineskip 13pt
Polarization asymmetry zero as a function of $Q^2/4m^2$, for
gluon virtuality $k^2 = 0$.}
\label{8370A02}
\end{figure}

Since the gluon entering the fusion process in sea-quark leptoproduction is a
constituent of the target hadron, it is always spacelike
$k^2 <0$, and  the sum rule is not exactly zero.    The
virtuality of the gluon can in fact be determined from the measurement of the
lepton and heavy pair kinematics. However, a zero crossing point still occurs
even for $k^2$ not zero. It can be calculated from the complete virtual photon-
virtual gluon polarization asymmetry:\cite{StTh,GR}
\begin{equation}
\begin{array}{clcr}
{\cal A}\big(x,Q^{2},K^{2}\big)=&-{\alpha _{s}\over 2\pi}{\sqrt
{1-{4m^{2}\over s}}\over 1-{4x^{2}K^{2}\over Q^{2}}}\left\{
\big(2x-1\big)\left(1-{2xK^{2}\over Q^{2}}\right)\right. \\[2ex]
&\times
\left[1-{1\over \sqrt {1-{4m^{2}\over s}}\sqrt {1-{4x^{2}K^{2}\over
Q^{2}}}}\ln \left({1+\sqrt {1-{4m^{2}\over s}}\sqrt {1-{4x^{2}K^{2}\over
Q^{2}}}\over 1-\sqrt {1-{4m^{2}\over s}}\sqrt {1-{4x^{2}K^{2}\over
Q^{2}}}}\right)\right] \\[4ex]
&+\left.\left(x-1+{xK^{2}\over
Q^{2}}\right){2\big(1-{4x^{2}K^{2}\over Q^{2}}\big)-{K^{2}\over
m^{2}}x\big(2x-1\big)\big(1-{2xK^{2}\over Q^{2}}\big)\over
\big(1-{4x^{2}K^{2}\over Q^{2}}\big)-{K^{2}\over m^{2}}x\big(x-1+{xK^{2}\over
Q^{2}}\big)}\right\},
\label{GG}
\end{array}
\end{equation}
where $K^2 = -k^2$, and
$s={Q^{2}(1-x)-K^{2}x\over x}$ is the invariant mass squared of the
photon-gluon system.
The result for the crossing point is shown in Fig. 3 as a function of $K^2/Q^2$,
for different
$Q^2/4m^2$ values.  The crossing point becomes
insensitive to the gluon virtuality for small values of $Q^2/4m^2$.

\vspace{0.5cm}
\begin{figure}[htb]
\begin{center}
\leavevmode {\epsfysize=6.8cm \epsfbox{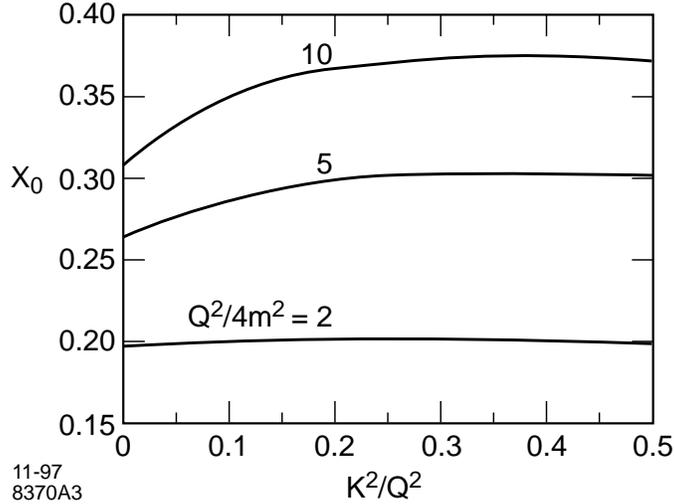}}
\end{center}
\caption[*]{\baselineskip 13pt
Polarization asymmetry zero as a function of $K^2/Q^2$, for
different values of $Q^2/4m^2$.}
\label{8370A03}
\end{figure}

The asymmetry zero and the underlying physics of the fusion process for
virtual gluons should be experimentally accessible. Consider the process
$\gamma^* q  \to c  \bar c q'.$ Clearly the virtuality of the exchanged
gluon can
be large since the quark emitting it can recoil at large momentum transfer. This
is the regime where the two heavy quarks and the emitting quark all appear as
final state jets at large transverse momentum. The inclusive deep inelastic
cross section is given by an integration
$\alpha_s(k^2) dk^2/k^2$ over the gluon virtuality corresponding to the
available span of transverse momentum of the emitting quark.  The region of
small virtuality dominates, with a logarithmic tail extending to the
kinematic limits.
If this three-jet final state is identified in an experiment, then one
would find zero contribution to the GEJ sum rule for events in which the
$q'$ has small transverse momentum compared to the mass of the heavy quark
$c$. The anomaly contribution $-(\alpha_s/2 \pi) \Delta G$ in the scaling
region
$Q^2 >> 4m_c^2$ thus derives from events where the gluon is off-shell with
$-k^2 \gsim m_c^2$, i.e., where the quark $q'$ recoils with transverse
momentum
of order $m_c$ or larger. There is no anomalous
contribution for $-k^2 \ll m_c^2$. The physics of the asymmetry zero thus
reflects the basic spin dynamics of the
$\gamma^* g \to c \bar{c}$ process, and it is connected to basic principles
underlying the gluon anomaly.

The asymmetry zero should also be a measurable effect in the case of the
polarized photon structure function. Consider $\gamma^* \gamma \to Q \bar{Q}$.
The asymmetry zero in the polarized photon-photon cross section has the same
expression as given before in terms of $Q^2$ and the heavy quark pair mass. The
target photon can be polarized using a back-scattered laser beam, and the
virtual photon polarization tracks with that of the scattered electron.

Since our analysis applies only
to the extrinsic contribution arising from photon-gluon fusion, the displacement
of the asymmetry zero from its canonical position can provide a test for the
presence of non-perturbative contributions to the charm photo- and
electroproduction cross sections. For example, the multiply-connected
(intrinsic) contributions\cite{IC}  to the charm structure function may give a
non-zero  contribution to the first moment of
$g_1^p(x)$ and displace the crossing point predicted by the photon-gluon fusion
contributions.   Although the intrinsic charm probability in the nucleon is of
order
$0.6\%$\cite{VHS}, the intrinsic contribution can dominate the charm structure
function at large momentum fraction $x_{BJ} > 0.2$ or near threshold. The
suppression of the  intrinsic bottom quarks is larger. On the other hand, there
could be substantial intrinsic effects in the case of strange
quarks\cite{StTh,Ma}.

We have thus demonstrated two unexpected features of the sea-quark
contributions to the polarized structure functions, a zero sum rule and a zero
crossing point, which can be traced directly to the dynamics of the perturbative
tree-graph gluon-splitting contributions.   In particular, we have shown
that the
Born contribution of massive quarks arising from the  photon-gluon fusion
subprocess gives zero contribution to the logarithmic integral over the
polarization asymmetry
$\int {d\nu \over \nu}\Delta
\sigma(\nu,Q^2)$ for any photon virtuality.
The vanishing of this integral in
the Bjorken scaling limit then implies a zero gluon-splitting Born contribution
to the Gourdin-Ellis-Jaffe sum rule for $\int^1_0 dx g_1^p(x,Q^2)$ from
massive sea quarks as long as the gluon virtuality can be
neglected. If the gluon virtuality $\langle
k^2\rangle$ is small compared to the  quark mass, the corrections will be of
order $\langle k^2\rangle /m^2$.  The displacement
of the asymmetry zero from its canonical position in photon energy provides a
measure of intrinsic and higher order sea quark contributions, as well as the
virtuality of the gluon in the photon-gluon fusion subprocess.

We thank Peter Bosted, Charles Hyde-Wright, Al Mueller,
and Tom Rizzo for helpful conversations.


\newpage

\begin{thebibliography}{99}

\bibitem{Ansel}
See for example:  M.  Anselmino,  A. Efremov, and E. Leader,
Phys. Rep. {\bf 261}, 1 (1995), and references therein.

\bibitem{Bj}
J. D. Bjorken,  Phys. Rev. {\bf D 1},  1376  (1971);
Phys. Rev. {\bf 148}, 1467  (1966).

\bibitem{CSR}
R. J. Crewther,  Phys. Rev. Lett. {\bf 28}, 1421 (1972);
S. J. Brodsky, G. T. Gabadadze, A. L. Kataev, and
H. J. Lu, Phys. Lett.  {\bf B372}, 133 (1996).

\bibitem{SMC97}
For the latest data see:
``Polarized Quark Distributions in the Nucleon from Semi-inclusive
Spin Asymmetries'', Spin Muon Collaboration, hep-ex/9711008.

\bibitem{Ma}
S. J. Brodsky and B.-Q. Ma,  Phys. Lett. {\bf B381} 317  (1996);
Phys. Lett. {\bf B392}, 452 (1997).

\bibitem{Ioffe}  V. D. Burkert and  B. L.  Ioffe,
J. Exp. Theor. Phys. {\bf 78},  619  (1994).
(Zh. Eksp. Teor. Fiz. {\bf 105},  1153  (1994)).

\bibitem{DHG}
{S.  D. Drell and A.  C. Hearn,  Phys.  Rev.  Lett. {\bf 16},  908
(1966);  S. Gerasimov, Yad. Fiz. {\bf 2},  598  (1965) (Sov.  J.  Nucl.
Phys. {\bf 2},  430 (1966)).}  For a recent review, see  S. D.
Bass, Mod. Phys. Lett. {\bf A12} 1051, (1997).

\bibitem{BS}
G. Altarelli,  N.~Cabibbo,  L. Maiani, Phys. Lett. {\bf 40B},  415  (1972);
S.  J.  Brodsky and  I. Schmidt,
Phys. Lett. {\bf B351},  344  (1995).

\bibitem{GEJ}
M.~Gourdin, Nucl. Phys. {\bf B 38},  418  (1972);
                J.~Ellis and R.~L.~Jaffe,
                Phys.~Rev.~{\bf D 9}, 1444  (1974);
                           {\bf 10}, 1669  (1974)(E).


\bibitem{Carlitz}
R. D. Carlitz, J.C. Collins, and
A. H. Mueller, Phys. Lett. {\bf B214}, 229 (1988).

\bibitem{Geoff}
G. T. Bodwin and J. Qiu,  Phys. Rev. {\bf D41},  2755  (1990).

\bibitem{footnote}
These matrix elements are often referred to as ``cross sections"
in the literature.

\bibitem{Cheng}
See for example: H.-Y. Cheng,  Int.  J. of  Mod.  Phys. {\bf
11},  5109  (1996).

\bibitem{Forte}
S. Forte, Phys. Lett. {\bf B224}, 189 (1989);  Nucl. Phys.
{\bf B331}, 1  (1990);  S. Forte and E. V. Shuryak, Nucl. Phys. {\bf
B357}, 153  (1991);  A. E. Dorokhov and N. I. Kochelev, Mod. Phys. Lett.
{\bf A5}, 55  (1990);  Phys. Lett. {\bf B245}, 609 (1990);  Phys. Lett.
{\bf B357}, 153 (1991); A. E. Dorokhov, N. I. Kochelev and Y. A. Zubov,
Int. J. Mod. Phys. {\bf A8}, 603 (1993).

\bibitem{BL}
G. P. Lepage and S. J. Brodsky, Phys. Rev {\bf D22}, 2157 (1980).

\bibitem{SLAC}
P. Bosted, private communication;  E143 Collaboration (K. Abe {\it et al.}),
Phys. Lett. {\bf B364}, 61 (1995).

\bibitem{HERMES}
C. Bloch, {\it et al.},
Nucl. Instrum. Methods {\bf  A354}, 437  (1995).

\bibitem{BRS}
S. J. Brodsky, T. Rizzo, and I. Schmidt,  Phys. Rev. {\bf D52},
4929 (1995).

\bibitem{StTh}
S. D. Bass, N. N. Nikolaev, and A. W. Thomas, University of Adelaide
Report ADP-133-T-80, 1990 (unpublished); S.  D. Bass,  B. L.  Ioffe,
N. N.  Nikolaev,
and A. W. Thomas,  J. Moscow
Phys.  Soc. {\bf 1}, 317  (1991);
W. Vogelsang,  Z. Phys. {\bf C50},  275
(1991); F. M. Steffens and A. W. Thomas, Phys. Rev. {\bf D53}, 1191 (1996).

\bibitem{GR}
The photoproduction cross section ($q^2=0$) is given in M. Gluck and E. Reya, Z.
Phys. {\bf C39}, 569 (1988).

\bibitem{IC}
S. J. Brodsky, P. Hoyer, C. Peterson, and N. Sakai,
Phys.  Lett. {\bf 93B}, 451 (1980).

\bibitem{VHS}
B. W. Harris, J. Smith, and R. Vogt, Nucl. Phys. {\bf B461}, 181
(1996).

\nonfrenchspacing
\end{thebibliography}
\end{document}